\documentclass{article}
\usepackage{spconf,amsmath,graphicx}
\usepackage[ruled, vlined, linesnumbered]{algorithm2e}
\usepackage{bbm}
\usepackage{float}
\usepackage{multicol}
\usepackage{makecell}
\DeclareMathOperator*{\argmin}{arg\,min}
\title{Enhancing into the codec: Noise Robust Speech Coding with Vector-Quantized Autoencoders}

\name{Jonah Casebeer\sthanks{Equal contribution. Work performed while at Amazon Web Services.}$^{\dagger}$ \qquad Vinjai Vale\footnotemark[1]$^{\flat}$  \qquad  Umut Isik$^{\sharp}$ \qquad}

\secondnm{\textit{Jean-Marc Valin}$^{\sharp}$ \qquad \textit{Ritwik Giri}$^{\sharp}$ \qquad \textit{Arvindh Krishnaswamy}$^{\sharp}$}

\address{$^{\dagger}$ University of Illinois at Urbana-Champaign, $^{\flat}$ Stanford University\\
          $^{\sharp}$ Amazon Web Services
}

\begin{document}
%
\maketitle

\begin{abstract}
Audio codecs based on discretized neural autoencoders have recently been developed and shown to provide significantly higher compression levels for comparable quality speech output. However, these models are tightly coupled with speech content, and produce unintended outputs in noisy conditions. Based on VQ-VAE autoencoders with WaveRNN decoders, we develop compressor-enhancer encoders and accompanying decoders, and show that they operate well in noisy conditions. We also observe that a compressor-enhancer model performs better on clean speech inputs than a compressor model trained only on clean speech. 
\end{abstract}

\section{Introduction}
\label{sec:intro}

Audio codecs compress speech signals by eliminating redundant and unnecessary information, with their design often leveraging extensive domain expertise to keep compression rates high, while keeping artifacts at a minimum. The most popular codecs, like the Opus codec in wideband mode, can produce high-quality speech compression at around $9$ kb/s \cite{valin2012definition}. Recently, there have been successful efforts in building learned codecs; starting with replacing the decoders with learned decoders for fixed encoders, which can operate as low as $2.4$ kb/s to $1.6$ kb/s \cite{kleijn2018wavenet, valin2019real}. These learned decoders leverage advances in speech synthesizing generative models such as WaveNet, WaveRNN, and LPCNet \cite{oord2016wavenet, lorenzo2018towards, valin2019lpcnet}. More recently, in \cite{garbacea2019low, zhen2020psychoacoustic}, the encoder and decoder were both learned in a joint fashion, by using quantized bottlenecks based on Vector-Quantized Variational Auto-Encoders~(VQ-VAE) \cite{van2017neural}, and soft-to-hard quantizers. The VQVAE based model improved at $1.6$ kb/s, on hand-designed encoders at $2.4$ kb/s.


VQ-VAE models are auto-encoders where latent vectors are quantized using a learned vector quantization scheme. These discrete representations have been shown to have a good inductive bias for speech and perform well on unsupervised acoustic unit discovery tasks \cite{chorowski2019unsupervised, chen2020unsupervised}. VQ-VAE models are apt for low-bitrate compression, as an input can be represented by a sequence of discrete codebook vector indices, while the location of codebook vectors can be hard-coded.

Fully learned codecs like \cite{garbacea2019low, zhen2020psychoacoustic} open new avenues for learning based compression and demonstrate strong results for compressing clean speech. However, they are tightly coupled with speech content and do not perform well under the noisy conditions a codec might encounter in the wild. In this work, we focus on making them robust to speech corrupted with noise. In \cite{lim2020robust}, this issue was addressed by training a noise-robust feature extractor based on the Siamese learning paradigm, and then training a WaveNet model conditioned on those features. Motivated by the performance of modern neural speech enhancers in removing unwanted noise and reverberation from audio signals \cite{xia2013speech, xu2014regression, weninger2015speech, isik2020poconet, valin2018hybrid}, we combine the learning based compression and learning based enhancement paradigms. We call the resulting paradigm "enhancing into the codec". The proposed model is based on VQ-VAE with a WaveRNN decoder, and, trained end-to-end as a speech enhancer, can simultaneously compress and enhance noisy speech signals, independent of speaker identity.
We refer to such a model as a compressor-enhancer: a model which jointly compresses and enhances speech.




We measure the performance of our models using Mean Opinion Scores (MOS) from a crowd-sourced study on Amazon Mechanical Turk. Across a range of compression rates and noise levels, we compare our model to a non-enhancing learned compressor both with and without additional enhancement preprocessing, as well as the LPCNet neural codec \cite{valin2019lpcnet}. We find that the proposed model performs well in noisy scenarios, compared to both non-enhancing codecs, as well as the composition, with comparable total compute cost, of a separate speech enhancer and a non-enhancing codec. We also find that the compressor-enhancer VQ-VAE performs significantly better than a clean-speech-trained VQ-VAE codec on clean speech inputs.


\section{Method}
\label{sec:method}

\begin{figure*}
    \centering
    \includegraphics[scale=0.75]{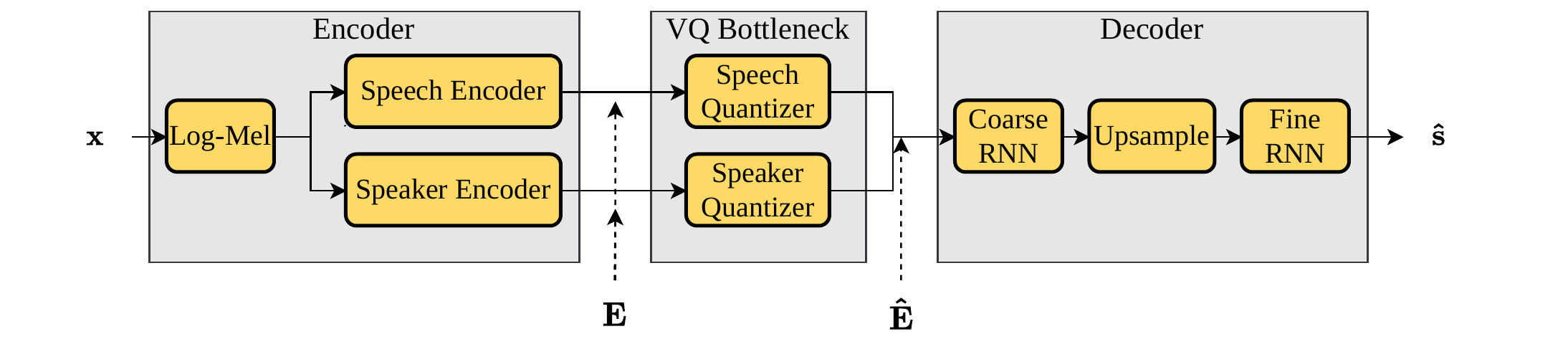}
    \vspace{-3mm}
    \caption{Block Diagram of the compressor-enhancer. The speech and speaker encoders are made up of several convolutional layers with batch normalization and ReLU. The VQ bottleneck has separate quantizers for the speech and speaker encodings. In our experiments the speech quantizer is made up of several codebooks of size 512 and the speaker quantizer is made up of a single codebook of size 512. The decoder is a WaveRNN based model and uses the quantized speech and speaker information first to reconstruct a coarse and then reconstruct a fine waveform. The waveform output is mu-law quantized.}
    \vspace{-2mm}
    \label{fig:model}
\end{figure*}

The task of speech enhancement is to recover a clean speech signal $\mathbf{s}$ from a noisy and possibly reverberant mixture $\mathbf{x} = \mathbf{s} * \mathbf{h} + \mathbf{n}$ where $\mathbf{n}$ represents some additive noise signal and $\mathbf{s} * \mathbf{h}$ represents the convolution of a room impulse response $\mathbf{h}$ with the speech signal. The goal of speech compression is to reconstruct a speech signal  $\mathbf{s}$ after encoding it to a smaller representation $\mathbf{\hat{E}}$ and decoding it to the reconstruction $\mathbf{\hat{s}}$. A neural compression model is composed of an encoder network $\mathcal{N}_e$, a coding step $\mathcal{C}$ and decoder network $\mathcal{N}_d$ which balance a trade-off between reconstruction fidelity and the size of $\mathbf{\hat{E}}$.

Thus, we define the joint compression-enhancement task. In the joint compression-enhancement task, the model receives a noisy input $\mathbf{x}$, which it encodes to a smaller representation $\mathbf{\hat{E}}$ and then decodes to an estimated clean and decompressed speech signal $\mathbf{\hat{s}}$. The full procedure is therefore

\vspace{-3mm}
\begin{equation}
    \mathbf{\hat{s}} = \mathcal{N}_d( \overbrace{\mathcal{C}(\underbrace{\mathcal{N}_e(\mathbf{x})}_{\mathbf{E}})}^{\mathbf{\hat{E}}}).
\end{equation}
\vspace{-4mm}

\subsection{Model}
The proposed autoencoder model~(Fig. \ref{fig:model}) is comprised of a convolutional encoder, a VQ-VAE bottleneck, and a recurrent decoder. These correspond to $\mathcal{N}_e$, $\mathcal{C}$, and $\mathcal{N}_d$ respectively. The model takes as input a 16 kHz noisy speech signal which is processed by the encoder and quantized by the bottleneck. The decoder autoregressively reconstructs the original 16 kHz waveform using the quantized speech and speaker encodings. As such, the encoder is encouraged to produce a compressed representation that gives the most information for the decoder to conditionally model the clean speech signal.  

\subsubsection{Encoder}
The encoder first computes a log-Mel representation and then applies a series of 1D convolutional layers, treating the mel bins as features. Each convolutional layer is followed by a batch normalization and then a ReLU non-linearity. The stride of the log-Mel representation and convolutional layers are selected to produce encodings at a rate of 50 Hz. The encoder also estimates one additional ``speaker embedding'' vector by performing a simple average across time over the output of a separate set of encoding layers. The output of theses steps is called $\mathbf{E}$. 

\subsubsection{Vector-Quantized Bottleneck}
The vector quantized bottleneck quantizes the outputs of the encoding layer using a set of codebooks. Where a separate codebook, constant over the entire input, is used to quantize the speaker embedding. We represent a codebook containing $K$ codes by $\mathbf{C} = \{\mathbf{c}_1, \cdots \mathbf{c}_K\}$. In the forward pass, the encoder outputs $\mathbf{E} = \{\mathbf{e}_1, \cdots \mathbf{e}_K\}$ are quantized by replacing each $\mathbf{e}_i$ with the closest $\mathbf{c}_j$ to get the quantized encoding $\mathbf{\hat{e}}_i$, where $j = \argmin_k ||\mathbf{e}_i - \mathbf{c}_k||_2^2$. Due to the non-differntiability of the $\argmin$ operation, VQ-VAE uses an additional two loss terms. The terms encourage each encoding $\mathbf{e}_i$ to be close to the selected $\mathbf{c}_j$ and for the each code $\mathbf{c}_j$ to minimize the quantization error incurred by any encodings that selected it. These are summarized below using the stopgradient operator $\operatorname{sg}$ which is identity at the forward pass but stops gradients in the reverse pass. In practice, we optimize the second term using an exponential moving average k-means. For additional details, see \cite{van2017neural}.

\begin{equation}
    \mathcal{L}_{\text{vq}} = \lambda ||\operatorname{sg}[\mathbf{\hat{E}}] - \mathbf{E}] ||_2^2 + ||\operatorname{sg}[\mathbf{E}] - \mathbf{\hat{E}}] ||_2^2.
\end{equation}

\subsubsection{Autoregressive Decoder}
We use an RNN based model to synthesize raw 16 kHz audio. The model, which is described in \cite{lorenzo2018towards}, contains two Gated Recurrent Units~(GRU) and two dense layers. We first concatenate the quantized speaker embedding to the quantized encoding and pass the resulting tensor through the first GRU. Then, we up-sample the GRU output to match the desired output length~(in raw audio samples) and pass the upsampled tensor through the second GRU, and two final dense layers. We apply softmax to the final dense layer and train the model to predict a distribution over 8-bit mu-law quantized values.

\subsubsection{Loss}
The final forward pass procedure is composed of passing raw noisy audio $\mathbf{x}$ to the encoder, quantizing the resulting encodings and speaker embedding, and running an autoregressive model to produce an estimated clean waveform $\mathbf{\hat{s}}$. The full loss function shown below is composed of $\mathcal{L}_{\text{vq}}$ and a cross-entropy term $\mathcal{L}_{\text{ce}}(\mathbf{s}, \mathbf{\hat{s}})$, which measures the KL-divergence between the predicted distribution and the one-hot value of the mu-law quantized clean speech $\mathbf{s}$.

\begin{equation}
    \mathcal{L} = \mathcal{L}_{\text{vq}} + \mathcal{L}_{\text{ce}}(\mathbf{s}, \mathbf{\hat{s}}).
\end{equation}




\section{Experiments and Results}
\label{sec:exp_res}
We train the Codec Only and Enhancing Codec models at two different kb/s and for the highest kb/s Enhancing Codec we also experiment with modifying the training setup to use higher SNR mixtures.


\subsection{Model Details}
The encoder first computes an $80$ bin log-mel representation with a hop-size of $10$ms and a window size of $250$ms on $16$kHz audio. These are passed to the speech encoder which has five convolutional layers each with $768$ filters. The first, second, fourth and fifth layers use a stride of $1$ and a kernel of size $3$. The third layer downsamples by using a stride of $2$ and a kernel of size $4$. The speaker encoder has an identical architecture but uses $64$ filters and omits the fifth layer. These are both passed to separate VQ bottlenecks which apply a linear layer with output size $64$ before quantizing. The speaker encoding for an entire input file is quantized using a single code from a $9$-bit codebook, while the speech encoding is quantized using two $9$-bit codebooks for the $.9\,\text{kb/s}$ model and three $9$-bit codebooks for the $1.35\,\text{kb/s}$  model. When several codebooks are used, each codebook uses its own linear layer and the resulting output quantizations are stacked. In our current implementation these steps are non-causal, but can easily be made causal or with custom look-ahead by using causal convolutions and adapting speaker-encodings over time. The WaveRNN model's first GRU which produces the coarse representation has $192$ hidden nodes, and its fine-representation GRU has $896$ hidden nodes. 

All VQ codebooks are trained using the exponential moving average technique from \cite{van2017neural}. We train the models with a batch size of 80 per GPU and a sample length of 1 second using the Adam optimizer on 8 NVidia V100 GPUs for 3 days.


\subsection{Datasets}

\subsubsection{Training}
\label{sec:data_train}
To generate a training mixture, we retrieve clean speech data from the LibriSpeech dataset in \cite{panayotov2015librispeech}, and noise data from AudioSet \cite{gemmeke2017audio}. When selecting noise clips from AudioSet, we avoid any clips with speech related tags. To increase the prevalence of challenging noise we sample noise clips with non-stationary noise more frequently. The noisy mixtures are created with a random SNR between $-5$ and $25$ dB. Finally, all room impulse responses are synthetically generated using the image-source method. For additional details consult \cite{isik2020poconet}.

\subsubsection{Evaluation}
To evaluate, we use the test mixtures from the VCTK dataset \cite{valentini2017noisy}. It contains mixtures with SNRs of $2.5$dB, $7.5$dB, $12.5$dB and $17.5$dB across a variety of speakers and noises. When evaluating our models on clean speech we use the clean speech samples from the VCTK test set.

\subsection{Subjective Quality Evaluation}
Since the compression based models in this paper resynthesize waveforms, their performance is not aptly measured by standard numerical metrics; we therefore measure model performance using a Mean Opinion Score (MOS) from a crowd-sourced study on Mechanical Turk that uses the P808 evaluation method \cite{P808}. 

\subsection{Results}
We compare the compressor-enhancer model with a compressor-only counterpart of identical architecture and size at bandwidths of $1.35\,\text{kb/s}$ and $0.9\,\text{kb/s}$ . The compressor only model is trained on the clean speech setup described in section~\ref{sec:data_train}. We also experiment with the combination of a speech enhancement model (RNNoise) \cite{valin2018hybrid} and the compression only model. We refer to these baselines as "Codec Only", and "Enhancement, then codec" respectively. As a final point of comparison we also evaluate a pre-trained LPCNet vocoder \cite{valin2019lpcnet}.

\subsubsection{MOS vs Bandwidth in Noisy Speech}
\begin{figure}
    \centering
    \includegraphics[scale=0.55]{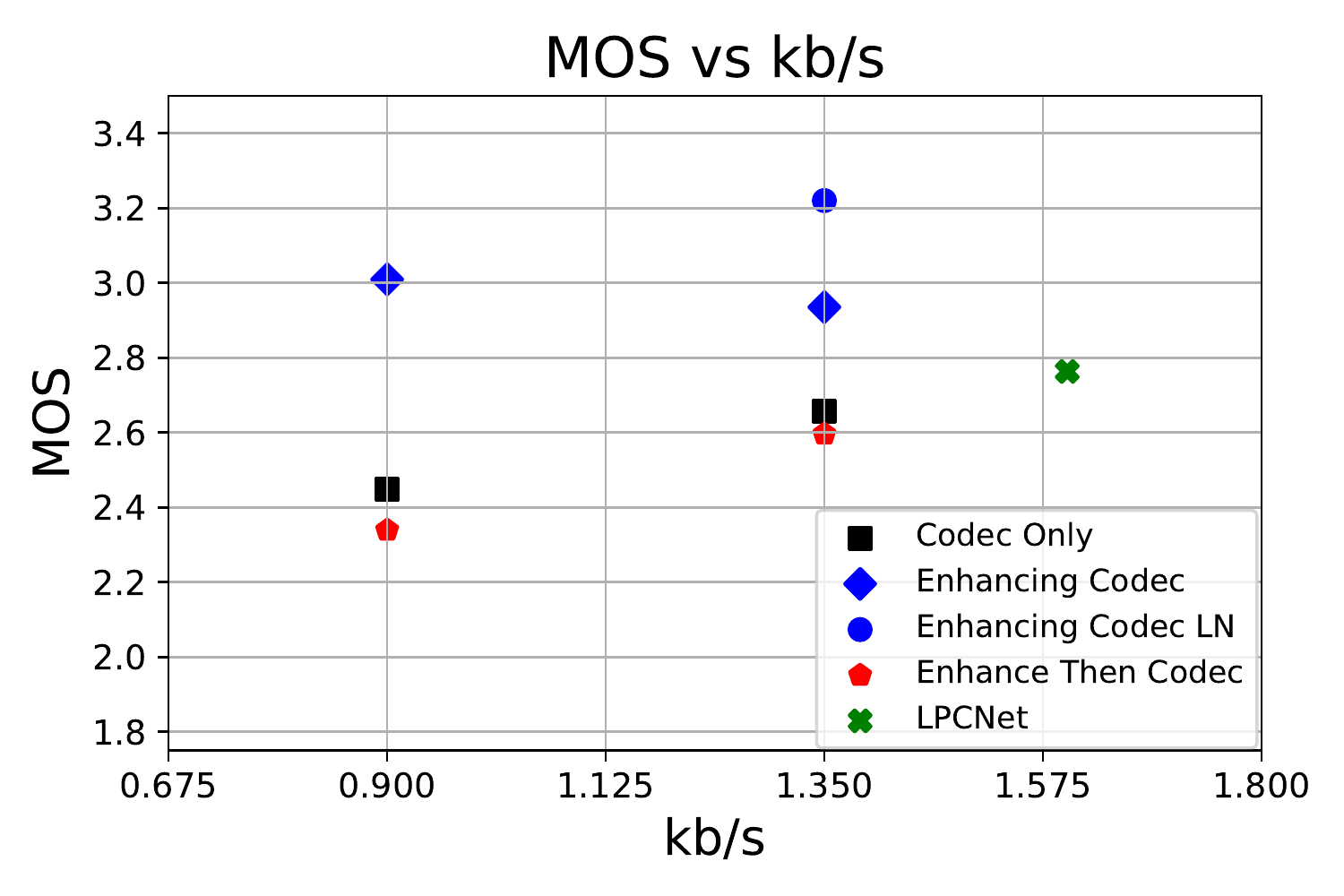}
    \vspace{-0.1in}
    \caption{MOS scores on the VCTK test set compared across models running at a range of kb/s. Our proposed joint compressor-enhancer models outperform the compression only baselines as well as the sequential enhancement-compression baselines. The LN suffix denotes a model trained on mixtures with lower noise content. The results are statistically significant with 95\% confidence intervals of approximately $.03$.}
    \label{fig:mos_v_bps}
\end{figure}

Fig.~\ref{fig:mos_v_bps} displays the MOS scores of the models at different bandwidths. Within the examined bandwidths, our enhancement-trained compression model~(blue diamonds and circle) scores about $.6$ MOS above its non-enhancing counterpart~(black square). Interestingly, sequential speech enhancement and compression scores worse than compression alone. We suspect this stems from the compression only model being susceptible to out of distribution errors. Finally, the LPCNet model serves as another comparison point for a compression only neural codec. The model denoted as "Enhancing Codec LN" is identical to "Enhancing Codec" in architecture but was trained on lower noise mixture with SNR ranging from $5$dB to $25$dB.

\subsubsection{Comparison of MOS across SNRs}
To see under what acoustic conditions enhancement-trained compression has the largest effect, we split MOS scores by SNR, and compare the $1.35\,\text{kb/s}$ versions of the three models discussed above. We display these results in Table \ref{Tab:diff_snr}. The MOS scores show that enhancement-trained compression compares favorably to both baselines across all SNRs. In the $2.5$dB scene the enhancing codec has a $.65$ to $.77$ lead over the baselines. This margin is reduced to $.38$ in the higher SNR scenes.

\begin{table}
    \begin{center}
            \vspace{-0.005in}
            \begin{tabular}{||c |c c c c||} 
            \hline
            Model \textbackslash SNR (dB) &  17.5 &  12.5 &  7.5 &  2.5 \\
            \hline\hline
             \thead{Codec Only}&  2.89&  2.79&  2.61&  2.34\\ 
            \hline
             \thead{Enhancing Codec LN}&  \textbf{3.27}&  \textbf{3.28}&  \textbf{3.22}&  \textbf{3.11}\\ 
            \hline
             \thead{Enhancement, then Codec}&  2.77&  2.64&  2.53&  2.46\\ 
            \hline
            \end{tabular}
    \end{center}
    \vspace{-0.1in}
    \caption{MOS comparison across SNRs on VCTK at $1.35\,\text{kb/s}$. The largest performance difference is in low SNR scenes.}
    \vspace{-0.1in}
    \label{Tab:diff_snr}
\end{table}

\subsubsection{Comparison of MOS on Clean Speech}
Observing the results at high-SNR, we also evaluated the MOS performance of enhancement-trained compression on clean speech. We compare the $1.35\,\text{kb/s}$ compression only model against our $1.35\,\text{kb/s}$ enhancement-trained compression model and display the results in Table \ref{Tab:clean}. We chose to omit the sequential enhancement then compression model since the speech is already clean. The enhancement-trained model outperforms the compression only model, leading us to suspect that training with noise helps the model learn a more robust bottleneck, and thus generalize better.

\begin{table}
    \begin{center}
            \begin{tabular}{||c | c||} 
            \hline
            Model&  Clean Speech MOS\\
            \hline\hline
             \thead{Codec Only}&  2.95\\ 
            \hline
             \thead{Enhancing Codec LN}&  \textbf{3.26}\\ 
            \hline
            \end{tabular}
    \end{center}
    \vspace{-0.1in}
    \caption{MOS comparison on VCTK clean speech at $1.35\,\text{kb/s}$. 95\% confidence intervals of $\approx .03$}
    \label{Tab:clean}
\end{table}

\subsection{Comparison with a two-stage approach}

We also attempted a joint compression and enhancement approach where we trained, first, a compression only model and then an encoder-only enhancement model trained to, given noisy speech, output the discretized latent representation of clean speech. With the goal being improvements to out-of-domain errors for speech enhancement models, as a two-stage approach would mean a fewer number of parameters needing to be trained on enhancement. However, we found that these models do not perform well, possibly because the clean-trained autoregressive decoder is too sensitive to out-of-domain inputs from the enhancer-encoder, making the simultaneous training of the decoder a key component of joint compression and enhancement.

\section{Conclusion}
\label{sec:conclusion}
In this work we presented a model that does joint compression and enhancement of a noisy speech signal using a VQ-VAE with a convolutional encoder and a WaveRNN decoder. 
Through a set of mean opinion score based experiments, we found that joint compression and enhancement performs better in the presence of noise, including in low SNR scenarios, than stand-alone compression; and also outperforms a sequential combination of speech enhancement and a compression only neural codec. We also found that enhancement training improves codec performance on clean speech signals. 








\bibliographystyle{IEEEbib}
\bibliography{strings,refs}

\begin{thebibliography}{10}

\bibitem{valin2012definition}
Jean-Marc Valin, Koen Vos, and Timothy Terriberry,
\newblock ``Definition of the opus audio codec,''
\newblock {\em IETF, September}, 2012.

\bibitem{kleijn2018wavenet}
W~Bastiaan Kleijn, Felicia~SC Lim, Alejandro Luebs, Jan Skoglund, Florian
  Stimberg, Quan Wang, and Thomas~C Walters,
\newblock ``Wavenet based low rate speech coding,''
\newblock in {\em 2018 IEEE International Conference on Acoustics, Speech and
  Signal Processing (ICASSP)}. IEEE, 2018, pp. 676--680.

\bibitem{valin2019real}
Jean-Marc Valin and Jan Skoglund,
\newblock ``A real-time wideband neural vocoder at 1.6 kb/s using lpcnet,''
\newblock {\em Proc. Interspeech 2019}, pp. 3406--3410, 2019.

\bibitem{oord2016wavenet}
Aaron van~den Oord, Sander Dieleman, Heiga Zen, Karen Simonyan, Oriol Vinyals,
  Alex Graves, Nal Kalchbrenner, Andrew Senior, and Koray Kavukcuoglu,
\newblock ``Wavenet: A generative model for raw audio,''
\newblock {\em arXiv preprint arXiv:1609.03499}, 2016.

\bibitem{lorenzo2018towards}
Jaime Lorenzo-Trueba, Thomas Drugman, Javier Latorre, Thomas Merritt, Bartosz
  Putrycz, Roberto Barra-Chicote, Alexis Moinet, and Vatsal Aggarwal,
\newblock ``Towards achieving robust universal neural vocoding,''
\newblock {\em arXiv preprint arXiv:1811.06292}, 2018.

\bibitem{valin2019lpcnet}
Jean-Marc Valin and Jan Skoglund,
\newblock ``Lpcnet: Improving neural speech synthesis through linear
  prediction,''
\newblock in {\em ICASSP 2019-2019 IEEE International Conference on Acoustics,
  Speech and Signal Processing (ICASSP)}. IEEE, 2019, pp. 5891--5895.

\bibitem{garbacea2019low}
Cristina G{\^a}rbacea, A{\"a}ron van~den Oord, Yazhe Li, Felicia~SC Lim,
  Alejandro Luebs, Oriol Vinyals, and Thomas~C Walters,
\newblock ``Low bit-rate speech coding with vq-vae and a wavenet decoder,''
\newblock in {\em ICASSP 2019-2019 IEEE International Conference on Acoustics,
  Speech and Signal Processing (ICASSP)}. IEEE, 2019, pp. 735--739.

\bibitem{zhen2020psychoacoustic}
Kai Zhen, Mi~Suk Lee, Jongmo Sung, Seungkwon Beack, and Minje Kim,
\newblock ``Psychoacoustic calibration of loss functions for efficient
  end-to-end neural audio coding,''
\newblock {\em IEEE Signal Processing Letters}, vol. 27, pp. 2159--2163, 2020.

\bibitem{van2017neural}
Aaron Van Den~Oord, Oriol Vinyals, et~al.,
\newblock ``Neural discrete representation learning,''
\newblock in {\em Advances in Neural Information Processing Systems}, 2017, pp.
  6306--6315.

\bibitem{chorowski2019unsupervised}
Jan Chorowski, Ron~J Weiss, Samy Bengio, and A{\"a}ron van~den Oord,
\newblock ``Unsupervised speech representation learning using wavenet
  autoencoders,''
\newblock {\em IEEE/ACM transactions on audio, speech, and language
  processing}, vol. 27, no. 12, pp. 2041--2053, 2019.

\bibitem{chen2020unsupervised}
Mingjie Chen and Thomas Hain,
\newblock ``Unsupervised acoustic unit representation learning for voice
  conversion using wavenet auto-encoders,''
\newblock {\em arXiv preprint arXiv:2008.06892}, 2020.

\bibitem{lim2020robust}
Felicia~SC Lim, W~Bastiaan Kleijn, Michael Chinen, and Jan Skoglund,
\newblock ``Robust low rate speech coding based on cloned networks and
  wavenet,''
\newblock in {\em ICASSP 2020-2020 IEEE International Conference on Acoustics,
  Speech and Signal Processing (ICASSP)}. IEEE, 2020, pp. 6769--6773.

\bibitem{xia2013speech}
Bingyin Xia and Changchun Bao,
\newblock ``Speech enhancement with weighted denoising auto-encoder.,''
\newblock in {\em Interspeech}, 2013, pp. 3444--3448.

\bibitem{xu2014regression}
Yong Xu, Jun Du, Li-Rong Dai, and Chin-Hui Lee,
\newblock ``A regression approach to speech enhancement based on deep neural
  networks,''
\newblock {\em IEEE/ACM Transactions on Audio, Speech, and Language
  Processing}, vol. 23, no. 1, pp. 7--19, 2014.

\bibitem{weninger2015speech}
Felix Weninger, Hakan Erdogan, Shinji Watanabe, Emmanuel Vincent, Jonathan
  Le~Roux, John~R Hershey, and Bj{\"o}rn Schuller,
\newblock ``Speech enhancement with lstm recurrent neural networks and its
  application to noise-robust asr,''
\newblock in {\em International Conference on Latent Variable Analysis and
  Signal Separation}. Springer, 2015, pp. 91--99.

\bibitem{isik2020poconet}
Umut Isik, Ritwik Giri, Neerad Phansalkar, Jean-Marc Valin, Karim Helwani, and
  Arvindh Krishnaswamy,
\newblock ``Poconet: Better speech enhancement with frequency-positional
  embeddings, semi-supervised conversational data, and biased loss,''
\newblock {\em arXiv preprint arXiv:2008.04470}, 2020.

\bibitem{valin2018hybrid}
Jean-Marc Valin,
\newblock ``A hybrid dsp/deep learning approach to real-time full-band speech
  enhancement,''
\newblock in {\em 2018 IEEE 20th International Workshop on Multimedia Signal
  Processing (MMSP)}. IEEE, 2018, pp. 1--5.

\bibitem{panayotov2015librispeech}
Vassil Panayotov, Guoguo Chen, Daniel Povey, and Sanjeev Khudanpur,
\newblock ``Librispeech: an asr corpus based on public domain audio books,''
\newblock in {\em 2015 IEEE International Conference on Acoustics, Speech and
  Signal Processing (ICASSP)}. IEEE, 2015, pp. 5206--5210.

\bibitem{gemmeke2017audio}
Jort~F Gemmeke, Daniel~PW Ellis, Dylan Freedman, Aren Jansen, Wade Lawrence,
  R~Channing Moore, Manoj Plakal, and Marvin Ritter,
\newblock ``Audio set: An ontology and human-labeled dataset for audio
  events,''
\newblock in {\em 2017 IEEE International Conference on Acoustics, Speech and
  Signal Processing (ICASSP)}. IEEE, 2017, pp. 776--780.

\bibitem{valentini2017noisy}
Cassia Valentini-Botinhao et~al.,
\newblock ``{Noisy speech database for training speech enhancement algorithms
  and TTS models},''
\newblock {\em University of Edinburgh. School of Informatics. Centre for
  Speech Technology Research (CSTR)}, 2017.

\bibitem{P808}
ITU-T,
\newblock {\em Recommendation {P}.808: Subjective evaluation of speech quality
  with a crowdsourcing approach}, 2018.

\end{thebibliography}

\end{document}